# Bounds on the Size of Small Depth Circuits for Approximating Majority

Kazuyuki Amano

Dept of Comput Sci, Gunma Univ, Tenjin 1-5-1, Kiryu, Gunma 376-8515, Japan
Email: amano@cs.gunma-u.ac.jp

**Abstract.** In this paper, we show that for every constant $0 < \epsilon < 1/2$ and for every constant $d \geq 2$, the minimum size of a depth $d$ Boolean circuit that $\epsilon$-approximates Majority function on $n$ variables is $\exp(\Theta(n^{1/(2d-2)}))$. The lower bound for every $d \geq 2$ and the upper bound for $d = 2$ have been previously shown by O'Donnell and Wimmer [ICALP'07], and the contribution of this paper is to give a matching upper bound for $d \geq 3$.

## 1 Introduction and Results

An investigation of the construction of small circuits for computing Majority function in various computational models has attracted many researchers. Interesting positive results (e.g., for comparator networks [3] or for monotone formulae [7]) as well as some negative results (e.g., for constant depth circuits [5]) have been obtained so far.

There also have been many researches on the construction of circuits to *approximate* the majority function. In this paper, we consider this problem in the model of constant depth circuits consisting of AND and OR gates with unbounded fan-in.

It seems that there are two major notions of "approximate-Majority" in this model. The first meaning of "approximate-Majority" is to compute a function that coincides with the majority function on every points such that the fraction of 1's in the inputs is at least 2/3, or is at most 1/3. The complexity of approximate-Majority of this notion is closely related to the complexity of probabilistic computations, and has been widely investigated (e.g., [1,2,8].)

The second meaning of "approximate-Majority", which we focus on in this paper, is to compute a function that disagrees with the majority function on at most $\epsilon$ fraction of all points. We call such a function an $\epsilon$-approximation of the majority function.

O'Donnell and Wimmer [6] first investigated this problem and obtained the following: (i) For every constant $0 < \epsilon < 1/2$ and every constant $d \geq 2$, any depth-$d$ circuit computing an $\epsilon$-approximation of the majority function on $n$ variables is $\exp(\Omega(n^{1/(2d-2)}))$, and (ii) When $d = 2$, this lower bound is optimal up to a constant factor in the exponent. The lower bound is proved by a combination of the argument based on the Håstad's switching lemma [5] (see also [4]) and the Kruskal-Katona Theorem developed in extremal set theory. The upper bound is proved by showing the existence of a DNF formula of size $\exp(O(\sqrt{n}))$ that $\epsilon$-approximates the majority function for every constant $0 < \epsilon < 1/2$.

In this paper, we extend their results and show that their lower bound is in fact optimal (again, up to a constant factor in the exponent) *for every constant d*. Precisely, we give a probabilistic construction of depth $d$ circuits of size $\exp(O(n^{1/(2d-2)}))$ that $\epsilon$-approximates the majority function, for every constant $0 < \epsilon < 1/2$ and for every constant $d \geq 3$. This is a main (and only) result of this paper.

The proof is a simple generalization of the technique used in a beautiful construction of $O(n^{5.3})$ size monotone formulas for the majority function by Valiant [7]. It should be noted that our circuit is monotone (i.e., without negated literals) and is formula (i.e., every gate has fan-out one).

The organization of the paper is as follows. In Section 2, we give some notations and definitions. In Section 3, we describe the framework of our construction. The proof of the main result is described in Section 4.

## 2 Notations and Definitions

For a binary string $x \in \{0,1\}^n$, $|x|$ denotes the number of 1's in $x$. The *majority function* on $n$ variables, which is denoted by $\mathsf{Maj}_n$, is a Boolean function defined by $\mathsf{Maj}_n(x) = 1$ iff $|x| \geq n/2$. For $0 < \epsilon < 1$, a Boolean function $f : \{0,1\}^n \to \{0,1\}$ is said an $\epsilon$-*approximation* for $\mathsf{Maj}_n$ if $f$ and $\mathsf{Maj}_n$ disagree on at most $\epsilon$ fraction of all inputs, i.e.,

$$\Pr_x[f(x) \neq \mathsf{Maj}_n(x)] \leq \epsilon,$$

where the probability is over the uniform distribution on $\{0,1\}^n$. For a set $S$, $\sharp S$ denotes the cardinality of $S$.

We consider single-output circuits that consists of unbounded fan-in AND and OR gates over the input literals, i.e., input variables and their negations. The *depth* of a circuit is the number of gates in a longest path from the output to an input. The *size* of a circuit is the number of AND and OR gates in it.

Throughout the paper, $e$ denotes the base of the natural logarithm.

## 3 Random Circuits

Let $W = (w_1, \ldots, w_d)$ be a $d$-tuple of integers such that $w_i \geq 2$ for every $1 \leq i \leq d$. The values of $w_i$ will be determined later. Define a sequence of random circuits $\mathbf{f}_0, \mathbf{f}_1, \ldots, \mathbf{f}_d$ on $X = \{x_1, \ldots, x_n\}$ recursively as follows:

1. $\mathbf{f}_0$ is a Boolean variable chosen uniformly from $X = \{x_1, \ldots, x_n\}$.
2. For odd $k$, $\mathbf{f}_k$ is an AND of $w_k$ independent copies of $\mathbf{f}_{k-1}$. For even $k$, $\mathbf{f}_k$ is an OR of $w_k$ independent copies of $\mathbf{f}_{k-1}$.

It is clear that $\mathbf{f}_d$ is a random circuit (in fact, formula) of depth $d$, where the bottom level consists of AND gates, and the fan-in of each gate at the $k$-th level is $w_k$. The number of gates in $\mathbf{f}_d$ is given by $1 + w_d + w_{d-1}w_d + \cdots + \prod_{k=2}^{d} w_k < 2\prod_{k=2}^{d} w_k$.

For $k = 0, \ldots, d$ and $i \in \{0, 1\}$, let $A_k^i(p) : [0, 1] \to [0, 1]$ be a function defined as follows:

$$A_0^1(p) = p, \quad \text{for every } p \in [0, 1]$$
$$A_k^1(p) = (A_{k-1}^1(p))^{w_k} \quad \text{for every odd } k, \text{and for every } p \in [0, 1]$$
$$A_k^0(p) = (A_{k-1}^0(p))^{w_k} \quad \text{for every even } k \text{ with } k \geq 2, \text{and for every } p \in [0, 1]$$
$$A_k^0(p) + A_k^1(p) = 1 \quad \text{for every } k, \text{and for every } p \in [0, 1].$$

When $\mathbf{f}_0$ gets one with probability $p$ then $\mathbf{f}_k$ outputs $i \in \{0, 1\}$ with probability $A_k^i(p)$. Note that $A_k^1(\cdot)$ ($A_k^0(\cdot)$, resp.) is monotonically increasing (decreasing, resp).

The following simple lemma relates the value of $A_k^i(\cdot)$'s with the size of $\epsilon$-approximator circuits for the majority function.

**Lemma 1.** *Suppose that, for a given $W = (w_1, \ldots, w_d)$, we have*

$$A_d^1\left(\frac{1}{2} - \frac{\epsilon}{\sqrt{n}}\right) \leq \epsilon, \tag{1}$$

*and*

$$A_d^0\left(\frac{1}{2} + \frac{\epsilon}{\sqrt{n}}\right) \leq \epsilon. \tag{2}$$

*Then there is a depth $d$ circuit of size less than $2\prod_{k=2}^{d} w_k$ that computes a $3\epsilon$-approximation for $\mathsf{Maj}_n$.*

**Proof** For every $x \in \{0, 1\}^n$ with $|x| \leq n/2 - \epsilon\sqrt{n}$, we have

$$\Pr_{\mathbf{f}_d}[\mathbf{f}_d(x) \neq \mathsf{Maj}_n(x)] \leq A_d^1\left(\frac{1}{2} - \frac{\epsilon}{\sqrt{n}}\right) \leq \epsilon,$$

since Eq. (1) and $A_d^1(\cdot)$ is monotonically increasing. Similarly, for every $x \in \{0,1\}^n$ with $|x| \geq n/2 + \epsilon\sqrt{n}$, we have

$$\Pr_{\mathbf{f}_d}[\mathbf{f}_d(x) \neq \mathsf{Maj}_n(x)] \leq A_d^0\left(\frac{1}{2} - \frac{\epsilon}{\sqrt{n}}\right) \leq \epsilon,$$

since Eq. (2) and $A_d^0(\cdot)$ is monotonically decreasing. These immediately implies that there is a depth $d$ circuit of size less than $2\prod_{k=2}^d w_k$ whose output disagrees with the majority function on at most

$$\epsilon\frac{2^n}{2} + \epsilon\frac{2^n}{2} + \sharp\{x \in \{0,1\}^n \mid n/2 - \epsilon\sqrt{n} < |x| < n/2 + \epsilon\sqrt{n}\}$$

inputs. The last term is upper bounded by

$$2\epsilon\sqrt{n}\binom{n}{n/2} \leq 2\epsilon\sqrt{n} \cdot \frac{2^n}{\sqrt{n}} = 2\epsilon \cdot 2^n,$$

where the first inequality follows from the Stirling formula. This completes the proof of the lemma. □

Note that, in this notation, a famous construction of $O(n^{5.3})$ size monotone formulae by Valiant [7] can be written as: $A_d^0(\alpha(1/2 + 1/n)) < 2^{-n}$ and $A_d^1(\alpha(1/2 - 1/n)) < 2^{-n}$ for $W = (2, 2, \ldots, 2)$ with $d \sim 5.3 \log_2 n$ and $\alpha = (\sqrt{5} - 1)/2$.

## 4 Bounds for Approximating Majority

In this section, we show our main theorem:

**Theorem 2.** *For every constant $0 < \epsilon < 1/2$ and for every constant $d \geq 3$, the majority function on $n$ variables can be $\epsilon$-approximated by a depth $d$ circuit of size $2^{O(n^{1/(2d-2)})}$. This is optimal up to a constant factor in the exponent.*

By Lemma 1, all we have to do is choose a suitable parameter $W = (w_1, \ldots, w_d)$ and verify that $A_d^0(1/2 + \epsilon/\sqrt{n}) \leq \epsilon$ and $A_d^1(1/2 - \epsilon/\sqrt{n}) \leq \epsilon$.

We first give a proof for the case $d = 3$ as an illustrative example in Section 4.1, and then give a proof for general cases in Section 4.2. The proof for general cases includes also the case $d = 3$, and so a reader can skip Section 4.1 and go directly to Section 4.2. The key ingredient of the proof is Lemmas 3 and 4 in Section 4.2.

### 4.1 Construction of Depth Three Circuits

We pick $W = (w_1, w_2, w_3)$ with $w_1 = \frac{1}{\epsilon} n^{1/4}$, $\tilde{w}_2 = 2^{w_1} w_1$, $w_2 = (\ln 2)\tilde{w}_2$, $\tilde{w}_3 = 2^{w_1}$ and $w_3 = (\ln 2)\tilde{w}_3$. The number of gates in a circuit that will be constructed is less than $2w_2 w_3 = 2(\ln 2)^2 (2^{w_1})^2 w_1 = 2^{O(n^{1/4}/\epsilon)}$.

Note that $A_1^1(1/2) = (1/2)^{w_1}$, $A_2^0(1/2) = (1 - (1/2)^{w_1})^{w_2} \sim (1/2)^{w_1}$ and $A_3^1(1/2) = (1 - (1/2)^{w_1})^{w_3} \sim 1/2$, which is a key property of our parameter. Put $p_h := 1/2 + \epsilon/\sqrt{n}$ and $p_\ell := 1/2 - \epsilon/\sqrt{n}$. Below we give a proof for $A_3^0(p_h) \leq \epsilon$ and $A_3^1(p_\ell) \leq \epsilon$, which is a bit long but uses only elementary calculations.

We first show that $A_3^0(p_h) \leq \epsilon$. By the definition, we have

$$A_2^0(p_h) = (1 - p_h^{w_1})^{w_2} = \left\{(1 - p_h^{w_1})^{(\ln 2 / p_h^{w_1})}\right\}^{p_h^{w_1} \tilde{w}_2} \leq \left(\frac{1}{2}\right)^{p_h^{w_1} \tilde{w}_2}. \quad (3)$$

Here we use the inequality $(1 - q)^{1/q} \leq 1/e$ for $q < 1$. The exponent in the last term of Eq. (3) is

$$p_h^{w_1} \tilde{w}_2 = \left(\frac{1}{2} + \frac{\epsilon}{\sqrt{n}}\right)^{w_1} \tilde{w}_2 = \tilde{w}_2 \left\{\left(\frac{1}{2}\right)^{w_1} \left(1 + \frac{2\epsilon}{\sqrt{n}}\right)^{w_1}\right\}$$

$$\geq \tilde{w}_2 \left(\frac{1}{2}\right)^{w_1} \left(1 + \frac{2\epsilon}{\sqrt{n}} w_1\right) = w_1 \left(1 + \frac{2}{n^{1/4}}\right). \quad (4)$$

Here we use the inequality $(1 + q)^r \geq 1 + qr$ for $q > 0$ and $r \geq 1$.

We proceed to the estimation of $A_3^0(p_h)$. Since $(1 - q)^r \geq 1 - qr$ for $q < 1$ and $r \geq 1$, we have

$$A_3^0(p_h) = 1 - (1 - A_2^0(p_h))^{w_3} \leq 1 - (1 - A_2^0(p_h) w_3) = A_2^0(p_h) w_3. \quad (5)$$

By plugging Eqs. (3) and (4) into Eq. (5), we have

$$A_3^0(p_h) \leq A_2^0(p_h) w_3 \leq (\ln 2) \left(\frac{1}{2}\right)^{w_1} \left(\frac{1}{2}\right)^{w_1 \frac{2}{n^{1/4}}} 2^{w_1}$$

$$= (\ln 2) \left(\frac{1}{2}\right)^{\frac{2}{\epsilon}} < (\ln 2) \frac{\epsilon}{2} < \epsilon,$$

where the second last inequality follows from $(1/2)^{2/\epsilon} < \epsilon/2$ which is equivalent to $(1/2) < (\epsilon/2)^{\epsilon/2}$. This holds since the minimum value of the function $q^q$ is $(1/e)^{1/e} \sim 0.6922 > (1/2)$.

We now turn to show $A_3^1(p_\ell) \leq \epsilon$, in which we should bound the value of $A_2^0$ from below.

$$A_2^0(p_\ell) = (1 - p_\ell^{w_1})^{w_2} = \left\{(1 - p_\ell^{w_1})^{(\ln 2/p_\ell^{w_1})}\right\}^{p_\ell^{w_1} \tilde{w}_2}$$

$$\geq \left\{(1 - p_\ell^{w_1})\frac{1}{e}\right\}^{(\ln 2) \cdot p_\ell^{w_1} \tilde{w}_2} > \left\{(1 - p_\ell^{w_1})\frac{1}{2}\right\}^{p_\ell^{w_1} \tilde{w}_2}. \quad (6)$$

We use $(1-1/q)^q \geq (1-1/q)(1/e)$ for $q > 1$ to derive the first inequality[1], and use $(1-q)^{\ln 2} > 1 - q$ to the second. The exponent in the last term is

$$p_\ell^{w_1} \tilde{w}_2 = \left(\frac{1}{2} - \frac{\epsilon}{\sqrt{n}}\right)^{w_1} \tilde{w}_2 = \tilde{w}_2 \left(\frac{1}{2}\right)^{w_1} \left(1 - \frac{2\epsilon}{\sqrt{n}}\right)^{w_1}$$

$$\leq w_1 \left(\frac{1}{2}\right)^{\frac{2\epsilon}{\sqrt{n}} \frac{1}{\ln 2} w_1} = w_1 \left(\frac{1}{2}\right)^{\frac{2}{\ln 2} \frac{1}{n^{1/4}}} \leq w_1 \left(1 - \frac{1}{\ln 2} \frac{1}{n^{1/4}}\right). \quad (7)$$

We use $(1 - 1/q)^q \leq 1/e$ for $q > 1$ to derive the first inequality, and use $(1/2)^{2q} \leq (1-q)$ for $q \leq 1/2$, which is equivalent to $(1/4) \leq (1-q)^{1/q}$, to derive the last inequality. By plugging Eq. (7) into Eq. (6), we can show that, for every sufficiently large $n$,

$$A_2^0(p) \geq \left(\frac{1}{2}\right)^{w_1(1 - 1/n^{1/4})}. \quad (8)$$

The proof of the above inequality is described in Appendix (Section 5.1).

We now proceed to the estimation of $A_3^1(p_\ell)$. Since $(1-q)^r \leq (1/e)^{qr}$ for $q \leq 1$ and $r \geq 0$, we have

$$A_3^1(p_\ell) = (1 - A_2^0(p_\ell))^{w_3} \leq \left(\frac{1}{2}\right)^{\tilde{w}_3 A_2^0(p_\ell)}.$$

In order to show $A_3^1(p_\ell) \leq \epsilon$, it is sufficient to show that $\tilde{w}_3 A_2^0(p_\ell) \geq \log_2(1/\epsilon)$. By Eq. (8), we have

$$\tilde{w}_3 A_2^0(p_\ell) \geq 2^{w_1} \left(\frac{1}{2}\right)^{w_1(1-1/n^{1/4})} = \left(\frac{1}{2}\right)^{-w_1/n^{1/4}} = 2^{1/\epsilon} > \log_2(1/\epsilon).$$

This completes the proof of Theorem 2 for $d = 3$.

---

[1] Proof: $(1 - 1/q)^q = (1 - 1/q)(1 - 1/q)^{q-1} = (1 - 1/q)(1 + 1/(q-1))^{-(q-1)} \geq (1 - 1/q)(1/e)$.

## 4.2 Construction for General Depths

We pick $W = (w_1, w_2, \ldots, w_d)$ such that

- $w_1 = (1/\epsilon)n^{1/(2d-2)}$,
- $\tilde{w}_k = 2^{w_1}w_1$ and $w_k = (\ln 2)\tilde{w}_k$ for $k = 2, \ldots, d-1$,
- $\tilde{w}_d = 2^{w_1}$ and $w_d = (\ln 2)\tilde{w}_d$.

As for the case $d = 3$, we choose parameters so that $A_1^1(1/2) = (1/2)^{w_1}$, $A_2^0(1/2) = (1-(1/2)^{w_1})^{w_2} \sim (1/2)^{w_1}$, $A_3^1(1/2) = (1-(1/2)^{w_1})^{w_3} \sim (1/2)^{w_1}$, and so on.

The following two lemmas are almost all that we need. The proof of these two lemmas is described in Appendix (Sections 5.2 and 5.3).

**Lemma 3.** *Let $w = (\ln 2)2^{w_1}w_1$. Suppose that $n$ is sufficiently large. Suppose also that*

$$A \geq \left(\frac{1}{2}\right)^{w_1}\left(1 + cn^{\frac{\alpha-d}{2(d-1)}}\right)$$

*for some $\alpha \in \{2, \ldots, d-1\}$ and some positive constant $c$. If $\alpha < d-1$, then*

$$(1-A)^w \leq \left(\frac{1}{2}\right)^{w_1}\left(1 - \frac{c}{2\epsilon}n^{\frac{(\alpha+1)-d}{2(d-1)}}\right).$$

*If $\alpha = d-1$, then*

$$(1-A)^w \leq \left(\frac{1}{2}\right)^{w_1}\left(\frac{1}{2}\right)^{\frac{c}{\epsilon}}.$$

**Lemma 4.** *Let $w = (\ln 2)2^{w_1}w_1$. Suppose that $n$ is sufficiently large. Suppose also that*

$$A \leq \left(\frac{1}{2}\right)^{w_1}\left(1 - cn^{\frac{\alpha-d}{2(d-1)}}\right),$$

*for some $\alpha \in \{2, \ldots, d-1\}$ and some positive constant $c$. If $\alpha < d-1$, then*

$$(1-A)^w \geq \left(\frac{1}{2}\right)^{w_1}\left(1 + \frac{c}{2\epsilon}n^{\frac{(\alpha+1)-d}{2(d-1)}}\right).$$

*If $\alpha = d-1$, then*

$$(1-A)^w \geq \left(\frac{1}{2}\right)^{w_1}\left(\frac{1}{2}\right)^{-\frac{c}{1.1\epsilon}}.$$

**Proof of Theorem 2** Let $W = (w_1, \ldots, w_d)$ be as described at the beginning of this subsection. The size of a circuit that will be constructed is less than $2\prod_{k=2}^{d} w_k = 2(\ln 2)^{d-1}(2^{w_1})^{d-1}(w_1)^{d-2} = 2^{O(n^{1/(2d-2)}/\epsilon)}$. Put $p_h := 1/2 + \epsilon/\sqrt{n}$ and $p_\ell := 1/2 - \epsilon/\sqrt{n}$. Below, we will show that $A_d^0(p_h) \leq \epsilon$ and $A_d^1(p_\ell) \leq \epsilon$.

We first show that $A_d^0(p_h) \leq \epsilon$. We start with

$$A_1^1(p_h) = \left(\frac{1}{2}\right)^{w_1}\left(1 + \frac{2\epsilon}{\sqrt{n}}\right)^{w_1}$$
$$\geq \left(\frac{1}{2}\right)^{w_1}\left(1 + 2n^{\frac{2-d}{2(d-1)}}\right) > \left(\frac{1}{2}\right)^{w_1}\left(1 + n^{\frac{2-d}{2(d-1)}}\right), \qquad (9)$$

where the first inequality follows from the inequality $(1+q)^r \geq 1 + qr$ for $q \geq 0$ and $r \geq 1$. We use Lemma 3 to get

$$A_2^0(p_h) = (1 - A_1^1(p_h))^{w_2} \leq \left(\frac{1}{2}\right)^{w_1}\left(1 - \frac{1}{2\epsilon}n^{\frac{3-d}{2(d-1)}}\right).$$

Then we use Lemma 4 to get

$$A_3^1(p_h) = (1 - A_2^0(p_h))^{w_3} \geq \left(\frac{1}{2}\right)^{w_1}\left(1 + \frac{1}{(2\epsilon)^2}n^{\frac{4-d}{2(d-1)}}\right).$$

By applying Lemmas 3 and 4 alternatively, we have

$$A_{d-2}^0(p_h) \leq \left(\frac{1}{2}\right)^{w_1}\left(1 - \frac{1}{(2\epsilon)^{d-3}}n^{\frac{-1}{2(d-1)}}\right) \qquad (10)$$

when $d$ is even, or we have

$$A_{d-2}^1(p_h) \geq \left(\frac{1}{2}\right)^{w_1}\left(1 + \frac{1}{(2\epsilon)^{d-3}}n^{\frac{-1}{2(d-1)}}\right) \qquad (11)$$

when $d$ is odd. Note that when $d = 3$ we have already obtained Eq.(11) as Eq.(9). By applying Lemma 3 or 4 once again, we obtain

$$A_{d-1}^1(p_h) \geq \left(\frac{1}{2}\right)^{w_1}\left(\frac{1}{2}\right)^{-\frac{1}{(1.1\epsilon)(2\epsilon)^{d-3}}}$$
$$\geq \left(\frac{1}{2}\right)^{w_1}\left(\frac{1}{2}\right)^{-\frac{1}{2\epsilon}} = \left(\frac{1}{2}\right)^{w_1} 2^{\frac{1}{2\epsilon}} \geq \left(\frac{1}{2}\right)^{w_1} \cdot \log_2(1/\epsilon) \quad (12)$$

when $d$ is even, and

$$A_{d-1}^0(p_h) \leq \left(\frac{1}{2}\right)^{w_1}\left(\frac{1}{2}\right)^{\frac{1}{(\epsilon)(2\epsilon)^{d-3}}} \leq \left(\frac{1}{2}\right)^{w_1}\left(\frac{1}{2}\right)^{\frac{1}{\epsilon}} \qquad (13)$$

when $d$ is odd. The case for even $d$ is finished by using Eq. (12):

$$A_d^0(p_h) = (1 - A_d^1(p_h))^{w_d}$$
$$\leq \left\{1 - \left(\frac{1}{2}\right)^{w_1} \log_2(1/\epsilon)\right\}^{(\ln 2)2^{w_1}} \leq \left(\frac{1}{2}\right)^{\log_2(1/\epsilon)} = \epsilon,$$

where the first inequality follows from the inequality $(1-q)^r \leq (1/e)^{qr}$ for $q \leq 1$ and $r \geq 0$. The case for odd $d$ is finished by using Eq. (13):

$$A_d^1(p_h) \geq \left\{1 - \left(\frac{1}{2}\right)^{w_1} \left(\frac{1}{2}\right)^{\frac{1}{\epsilon}}\right\}^{(\ln 2)2^{w_1}}$$
$$\geq 1 - (\ln 2)\left(\frac{1}{2}\right)^{\frac{1}{\epsilon}} > 1 - (\ln 2)\epsilon > 1 - \epsilon.$$

Here we use the inequality $(1-q)^r \geq 1 - qr$ for $q \leq 1$ and $r \geq 1$ to derive the first inequality, and use $(1/2)^{(1/q)} < q$, which is equivalent to $(1/2) < q^q$, to the second. This holds since the minimum value of the function $q^q$ is $(1/e)^{(1/e)} \sim 0.6922$.

We now turn to show $A_d^1(p_\ell) \leq \epsilon$. The proof is almost analogous to the proof for $A_d^0(p_h) \leq \epsilon$. The "base" is

$$A_1^1(p_\ell) = \left(\frac{1}{2}\right)^{w_1} \left(1 - \frac{2\epsilon}{\sqrt{n}}\right)^{w_1}$$
$$\leq \left(\frac{1}{2}\right)^{w_1} \left(\frac{1}{2}\right)^{\frac{2}{\ln 2} n^{\frac{2-d}{2(d-1)}}}$$
$$\leq \left(\frac{1}{2}\right)^{w_1} \left(1 - \frac{1}{\ln 2} n^{\frac{2-d}{2(d-1)}}\right) < \left(\frac{1}{2}\right)^{w_1} \left(1 - n^{\frac{2-d}{2(d-1)}}\right), \quad (14)$$

where the first inequality follows from the inequality $(1-q)^r \leq (1/e)^{qr}$ for $q \leq 1$ and $r \geq 0$, and the second inequality follows from the inequality $(1/2)^{2q} \leq 1 - q$ for $q \leq 1/2$, which is equivalent to $(1/4) \leq (1-q)^{1/q}$. By applying Lemmas 3 and 4 alternatively, we have

$$A_{d-2}^1(p_\ell) \leq \left(\frac{1}{2}\right)^{w_1} \left(1 - \frac{1}{(2\epsilon)^{d-3}} n^{\frac{-1}{2(d-1)}}\right),$$

when $d$ is odd (note again that when $d = 3$, we have already obtained this as Eq.(14)), or we have

$$A_{d-2}^0(p_\ell) \geq \left(\frac{1}{2}\right)^{w_1} \left(1 + \frac{1}{(2\epsilon)^{d-3}} n^{\frac{-1}{2(d-1)}}\right)$$

when $d$ is even. These inequalities are identical to Eqs. (10) and (11) if we swap $p_h$ and $p_\ell$, "odd" and "even", and the role of 0 and 1. This immediately implies the desired bound, i.e., $A_d^1(p_\ell) \leq \epsilon$, since we have shown $A_d^0(p_h) \leq \epsilon$ from Eqs. (10) and (11). $\square$

## 5 Appendix

### 5.1 Proof of Eq. (8)

What we want to show is

$$\left\{(1-p_\ell^{w_1})\frac{1}{2}\right\}^{w_1(1-\frac{1}{\ln 2}\frac{1}{n^{1/4}})} \geq \left(\frac{1}{2}\right)^{w_1(1-\frac{1}{n^{1/4}})}.$$

This is equivalent to

$$1 - p_\ell^{w_1} \geq \left(\frac{1}{2}\right)^{\frac{1-\ln 2}{(\ln 2)n^{1/4}-1}}. \tag{15}$$

Since $1 - q \geq (1/2)^{2q}$ for $q \leq 1/2$, we have

$$1 - \Theta\left(\frac{1}{n^{1/4}}\right) = 1 - \frac{1}{2} \cdot \frac{1-\ln 2}{(\ln 2)n^{1/4}-1} \geq \text{RHS of Eq. (15)}.$$

Since $p_\ell^{w_1}$ is exponentially small in $n$, i.e., $p_\ell^{w_1} = O(1/2^{n^{1/4}}) = o(1/n^{1/4})$, Eq. (15) holds for sufficiently large $n$. $\square$

## 5.2 Proof of Lemma 3

Since $(1-q)^r \leq (1/e)^{qr}$ for $q \leq 1$ and $r \geq 0$, we have

$$(1-A)^w = (1-A)^{(\ln 2)2^{w_1} w_1} \leq \left(\frac{1}{2}\right)^{A \cdot 2^{w_1} w_1}$$

$$\leq \left(\frac{1}{2}\right)^{w_1(1+cn^{\frac{\alpha-d}{2(d-1)}})}$$

$$= \left(\frac{1}{2}\right)^{w_1} \left(\frac{1}{2}\right)^{c \cdot w_1 n^{\frac{\alpha-d}{2(d-1)}}}$$

$$= \left(\frac{1}{2}\right)^{w_1} \left(\frac{1}{2}\right)^{\frac{c}{\epsilon}n^{\frac{(\alpha+1)-d}{2(d-1)}}}.$$

This completes the proof for $\alpha = d - 1$. If $\alpha < d - 1$, then the exponent of the last term converges to 0 as $n \to \infty$. Hence, we use the inequality $(1/2)^q \leq (1 - q/2)$ for $q \leq 1$, which is equivalent to $(1/4) \leq (1 - q/2)^{2/q}$, to show

$$(1-A)^w \leq \left(\frac{1}{2}\right)^{w_1} \left(1 - \frac{c}{2\epsilon}n^{\frac{(\alpha+1)-d}{2(d-1)}}\right) \quad \text{(for sufficiently large } n\text{)},$$

which completes the proof of the lemma. □

## 5.3 Proof of Lemma 4

By using the inequality $(1 - 1/q)^q \geq (1 - 1/q)(1/e)$ for $q > 1$ (whose proof is in the footnote in Section 4.1), we have

$$(1-A)^w = (1-A)^{(\ln 2)2^{w_1} w_1} \geq \left\{(1-A)\left(\frac{1}{2}\right)\right\}^{A \cdot 2^{w_1} w_1}$$

$$\geq \left\{(1-A)\left(\frac{1}{2}\right)\right\}^{w_1(1-cn^{\frac{\alpha-d}{2(d-1)}})}$$

$$\geq \left(\frac{1}{2}\right)^{w_1(1-\frac{c}{1.1}n^{\frac{\alpha-d}{2(d-1)}})} \quad \text{(for sufficiently large } n\text{)}$$

$$= \left(\frac{1}{2}\right)^{w_1} \left(\frac{1}{2}\right)^{-\frac{c}{1.1\epsilon}n^{\frac{(\alpha+1)-d}{2(d-1)}}},$$

where the third inequality can be derived by a similar calculation as the proof of Eq. (8) in Section 5.1. This completes the proof for the case $\alpha = d - 1$. When $\alpha < d - 1$, the exponent of the last term converges to 0 as $n \to \infty$. Hence, we can use the inequality $2^q \geq (1 + (\ln 2)q)$ for $q < 1$, which is equivalent to $e \geq (1+q)^{1/q}$, to show

$$(1-A)^w \geq \left(\frac{1}{2}\right)^{w_1} \left\{1 + \frac{(\ln 2)c}{1.1\epsilon} n^{\frac{(\alpha+1)-d}{2(d-1)}}\right\} \quad \text{(for sufficiently large } n\text{)}$$

$$> \left(\frac{1}{2}\right)^{w_1} \left\{1 + \frac{c}{2\epsilon} n^{\frac{(\alpha+1)-d}{2(d-1)}}\right\}.$$

This completes the proof of the lemma. □